\documentclass[]{spie}  
\usepackage[]{graphicx, url}

\title{On-sky low order non-common path correction of the GPI
  Calibration Unit}

\author{Markus Hartung\supit{a}, Bruce Macintosh\supit{b}, Paul
  Langlois\supit{a}, Naru Sadakuni\supit{a}, Don Gavel\supit{c}, J. Kent Wallace\supit{d}, Dave
  Palmer\supit{e}, Lisa Poyneer\supit{e}, Dmitry Savransky\supit{f},
  Sandrine Thomas\supit{g}, Darren \\ Dillon\supit{c}, Jennifer Dunn\supit{h},  Pascal Hibon\supit{a}, Fredrik Rantakyro\supit{a}, and Stephen Goodsell\supit{a}
  \skiplinehalf
  \supit{a}Gemini Observatory, Casilla 603, La Serena, Chile;
  \\ \supit{b}Kavli Institute for Particle Astrophysics and Cosmology, Stanford University, Stanford, CA, United States
  \\ \supit{c}University of California Observatories/Lick Observatory, University of California, Santa Cruz, United States
  \\ \supit{d}Jet Propulsion Laboratory/California Institute of Technology, Pasadena, United States
  \\ \supit{e}Lawrence Livermore National Laboratory, Livermore, United States
  \\ \supit{f}Sibley School of Mechanical and Aerospace Engineering, Cornell University, Ithaca, NY 14853, United States
  \\ \supit{g}NASA Ames, Mountain View, Unites States
  \\ \supit{h}National Research Council of Canada Herzberg, Victoria, Canada}

\authorinfo{Send correspondence to Markus Hartung: mhartung@gemini.edu, Telephone: +56-51-2205 664}

\begin{document} 
  \maketitle 

\begin{abstract}
The Gemini Planet Imager (GPI) entered on-sky commissioning phase, and
had its First Light at the Gemini South telescope in November
2013. Meanwhile, the fast loops for atmospheric correction of the
Extreme Adaptive Optics (XAO) system have been closed on many dozen
stars at different magnitudes (I=4-8), elevation angles and a variety
of seeing conditions, and a stable loop performance was achieved from
the beginning. Ultimate contrast performance requires a very low
residual wavefront error (design goal 60 nm RMS), and optimization of
the planet finding instrument on different ends has just begun to
deepen and widen its dark hole region. Laboratory raw contrast
benchmarks are in the order of $10^{-6}$ or smaller. In the telescope
environment and in standard operations new challenges are faced
(changing gravity, temperature, vibrations) that are tackled by a
variety of techniques such as Kalman filtering, open-loop models to
keep alignment to within 5 mas, speckle nulling, and a calibration
unit (CAL). The CAL unit was especially designed by the Jet Propulsion
Laboratory to control slowly varying wavefront errors at the focal
plane of the apodized Lyot coronagraph by the means of two wavefront
sensors: 1) a 7x7 low order Shack-Hartmann SH wavefront sensor
(LOWFS), and 2) a special Mach-Zehnder interferometer for mid-order
spatial frequencies (HOWFS) - atypical in that the beam is split in
the focal plane via a pinhole but recombined in the pupil plane with a
beamsplitter. The original design goal aimed for sensing and
correcting on a level of a few nm which is extremely challenging in a
telescope environment. This paper focuses on non-common path low order
wavefront correction as achieved through the CAL unit on sky. We will
present the obtained results as well as explain challenges that we are
facing.
\end{abstract}


\keywords{High contrast imaging, non-common path aberrations}

\section{INTRODUCTION}
\label{sect:intro}  
In high contrast imaging systems the control of non-common path
aberrations (NCPA) is another key element for contrast
performance. This has been addressed for GPI\cite{macintosh_pnas_2014}
with the design of a dedicated CAL unit.  The HOWFS was specified to a
wavefront error of 1 nm RMS over spatial frequencies of 3-22
cycles/pupil) and the low-order wavefront sensor (LOWFS) to 5 nm RMS
for spatial frequencies $<3$ cycles/pupil.  In the laboratory the CAL
unit was demonstrated to reach a performance of $< 5$\,nm RMS for the
LOWFS and $\approx 5$\,nm RMS for the HOWFS, but this was not done
with the benefit of the full AO system of GPI but using a set-up with
a phase aberration screen.\cite{wallace_2010}. Not surprisingly, the
commissioning of all functionalities of this subsystem poses one of
our greatest challenges since we are aiming to reach a precision of a
few nanometers in the telescope environment that is subject to a
changing gravity vector, thermal drifts, vibrations, etc. We remind
the reader that there are also other techniques to sense and control
NCPA such as phase diversity and speckle nulling, and GPI indeed makes
use of speckle nulling to further lower the noise floor of the ``dark
hole''\cite{savransky_2012}. Although these other techniques exist,
the sensing must occur on time-scales over which the speckles evolve.
Speckle decorrelation times on other telescopes are on the order of 30
to 60 seconds.\cite{hinkley_2009} Therefore interrupting science
observations on these timescales is impractical, and the baseline
method of contemporaneous calibration wavefront sensing and science
observations was decided upon.

In this paper we describe the status and share our challenges of the
on-sky commissioning of the LOWFS which has started recently. Due to
internal vibrations caused by the cryocoolers the commissioning of the
Mach-Zehnder type interferometer of the CAL has been
postponed. Currently, we are in the process of investigating to
upgrade our cryocoolers with an active vibration cancellation
system.\footnote{Sunpower recently advertised its new ACS (Active
  Cancellation System) for the GT and other models. Our current
  Sunpower GT coolers are damped passively only with tuned vibration
  absorbers.\cite{hartung_2014}}

After GPI had its First Light in November 2013 the commissioning team
concentrated on basic and high priority tasks such as loop stability,
integration into Gemini science operations\cite{rantakyro_2014},
vibration mitigation\cite{hartung_2014} and AO controller optimization
\cite{poyneer_2014}. In the most recent commissioning run (4th run,
May 2014) with all core functionalities working and verified, we could
shift part of our resources to continue the commissioning of GPI's CAL
unit.


   \begin{figure}
   \begin{center}
   \begin{tabular}{c}
   \includegraphics[angle=0,width=0.7\textwidth]{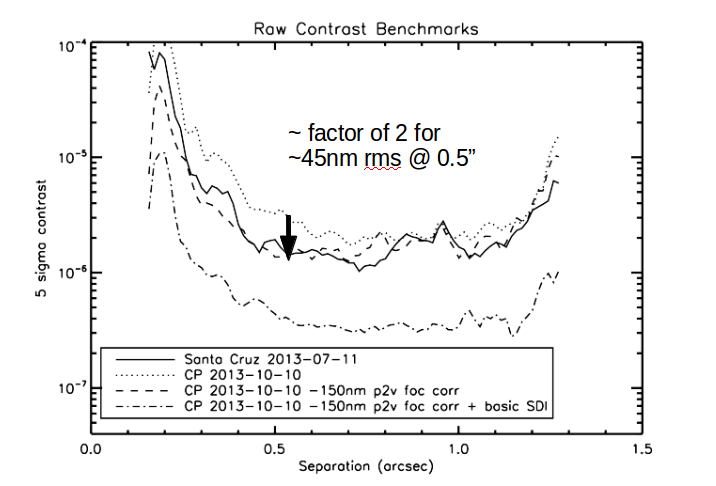}
   \end{tabular}
   \end{center}
   \caption[example] 
   { \label{fig:contrast} A defocus as small as 45\,nm RMS cost us
     approximately a factor of two in contrast at 0.5''. These data
     were taken during laboratory commissioning phase on Cerro Pachon
     while the focus was verified after shipment in a knife-edge
     experiment\cite{hartung_2013}.}
   \end{figure}

During the Assembly \& Testing (A\&T) stage\cite{hartung_2013} at the
University of California, Santa Cruz we established the nominal focus
using a knife edge experiment at the Focal Plan Mask (FPM), and the
corresponding wavefront sensor centroid offsets define the zero
point. We tested effects that potentially can create focus offsets. A
chromatic effect can make calibrations more challenging, and partly
illuminated subapertures might be more susceptible to instrument
misalignments or other effects. When applying the LOWFS measured and
reconstructed wavefront, a careful eye is required on the interplay
between tip/tilt and coma. Furthermore, the spherical has a larger
measurement error than the other Zernikes due to our geometry. With a
sampling of 7 subapertures over the pupil and the central obscuration
the central cap of the Zernike mode is blended out and this results
into a large fitting error.

Based on our A\&T experiences we chose a low-risk approach for on-sky
commissioning of the CAL unit and decided to reduce the complexity and
ignore the higher order modes for the first steps of on-sky
commissioning. Also, we expect to get the highest immediate gain from
assuring a correct non-common path focus. A defocus as small as 45\,nm
rm cost us approximately a factor of two in contrast at 0.5'' as we
demonstrated during commissioning at Cerro Pachon\cite{hartung_2013}
(Fig.~\ref{fig:contrast}). Therefore, this mode is prioritized
particularly in the light of moving GPI into Gemini standard science
operations by the end of this year. As soon as the off-load interfaces
are verified higher modes can be safely added.


   \begin{figure}
   \begin{center}
   \begin{tabular}{c}
   \includegraphics[angle=0,width=1.0\textwidth]{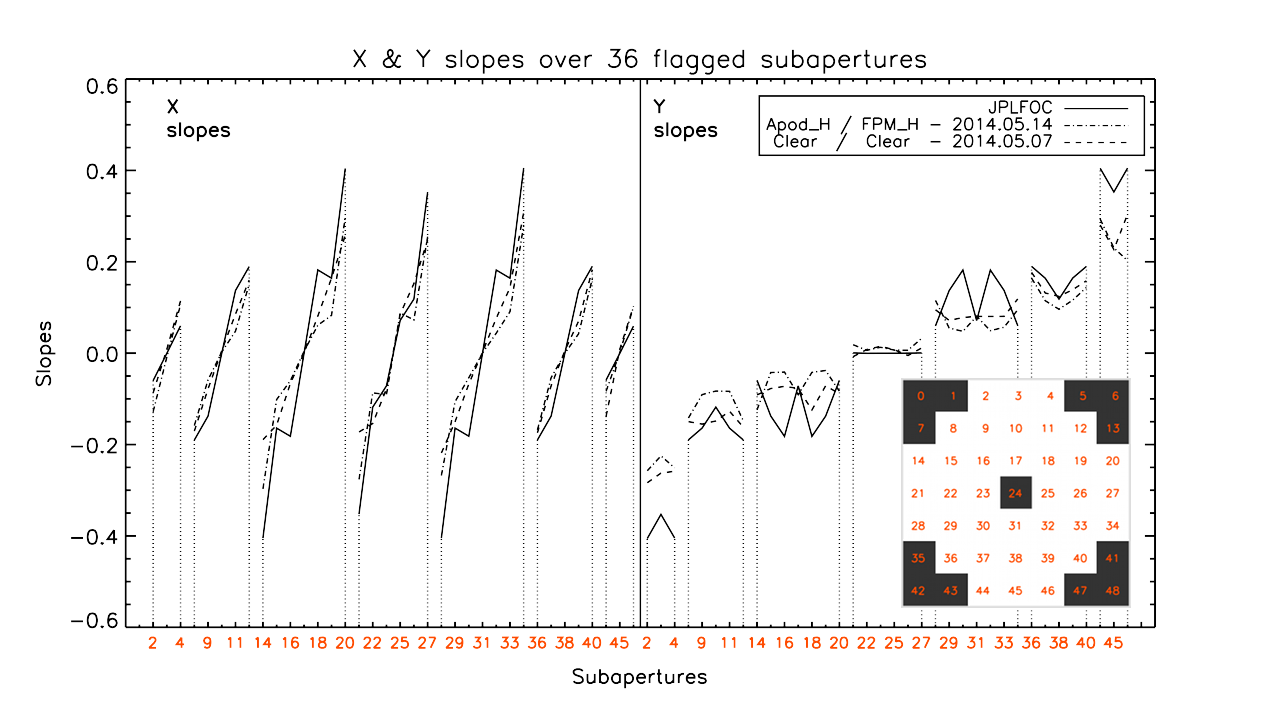}
   \end{tabular}
   \end{center}
   \caption[example] { \label{fig:slopes_compare} A 200 nm RMS focus
     offset was applied with the AO system in closed loop via
     reference centroids. The dash-dotted line displays the measured
     LOWFS slopes for this test wavefront with the apodizer (Apod\_H)
     inserted - a typical coronagraphic set up. The simple dashed
     line measures the same test wavefront but with the apodizer mask
     removed (Clear).  The continuous line represents the values of
     the focus row from the original synthetic reconstructor matrix,
     scaled to match the measured slopes.}
   \end{figure}

\section{Techniques and laboratory tests}
\subsection{LOWFS wavefront measurement and offload} \label{sect:offload}
The wavefront reconstruction of the LOWFS is based on a standard
vector matrix multiplier (VMM) approach.  In the original design, this
phase reconstruction matrix was derived synthetically (the slopes
corresponding to the different Zernike modes were simulated using an
appropriate geometry for the pupil, subapertures, the beam, and
detector pixel size). According to the sampling of 7 subapertures
across the pupil, the default size of the reconstruction matrix is
14x72, containing 14 Zernike coefficients (rows) and 72 slopes
(columns; first all x slopes, then all y slopes).  The inlet in
Fig.~\ref{fig:slopes_compare} shows the geometry and the valid
subapertures. There are three invalid subapertures at the four edges
and one in the center (the central subaperture is obscured by the
secondary mirror) which leaves us with 72 valid slopes.
The 14 Zernike coefficients representing the wavefront as measured
by the LOWFS (output of the VMM) are converted into a 48x48 phase
grid. To compensate the measured NCPA this phase grid is rendered to
the AO system. The tip/tilt offload loop is commissioned already, and the
higher-order off-load loop is currently in its commissioning phase.

\subsection{Laboratory tests}


  \begin{figure}
   \begin{center}
   \begin{tabular}{c}
   \includegraphics[angle=0,width=1.0\textwidth]{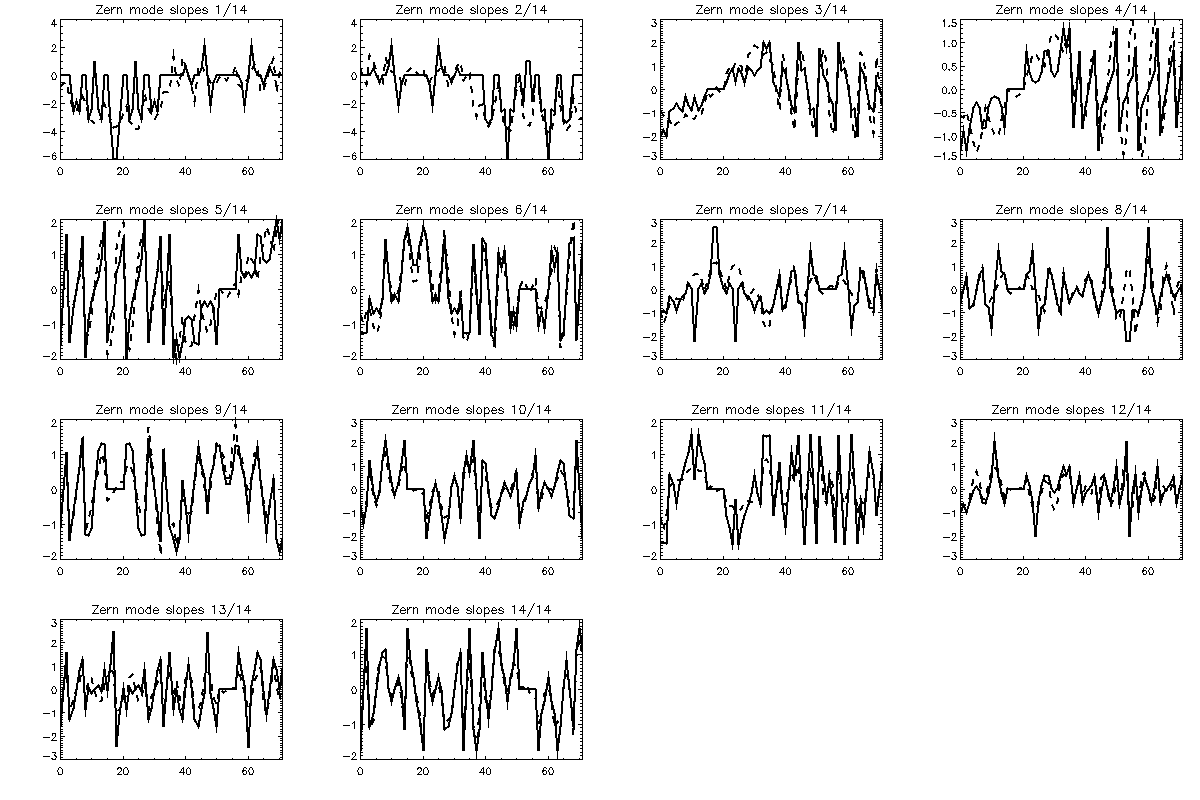}
   \end{tabular}
   \end{center}
   \caption[example] 
   { \label{fig:compare_mat} Comparison of the slopes for all modes of
     the synthetic matrix (original design) with the slopes (dashed
     lines) measured by pushing the corresponding Zernike modes
     (without focal plane mask).}
   \end{figure}

One reason for the discrepancy of the LOWFS focus measurements
compared to knife-edge measurements appeared to be the partly
illuminated subapertures. Indeed, temporarily ``flagging out'' all
partly illuminated subapertures significantly improved the accuracy of
absolute focus measurements (a low sensitivity for the spherical mode
also contributes to these discrepancies).

Inspection and additional measurements on CAL/LOWFS alignment
(collimation, lenslet, foci) did not reveal any significant anomalies
with respect to the design - still we might not have caught all
potential systematic discrepancies. Therefore, we have been
investigating to replace the synthetic reconstruction matrix with a
matrix based on measured slopes using the GPI's internal calibration
source (ASU - Artificial Star Unit). Fig.~\ref{fig:compare_mat} shows
a comparison of the slopes for all Zernike modes of the synthetic
matrix from the original design with the slopes that were measured by
pushing the required Zernike modes using the AO system.  Any
instrumental effects that cannot be simulated such as systematic
offsets or misalignments should be included in this measured matrix. A
potential systematic error in the original reconstruction might be
from the diameter of the central obstruction of the telescope which
was improperly interpreted in the design phase, and thus improperly
included in the as-delivered synthetic reconstruction matrix. We
remind the reader that since we are comparing the slopes of the
reconstruction matrices, the measured forward matrix was inverted via
a standard singular value decomposition (SVD).


Fig.~\ref{fig:slopes_compare} shows a direct comparison of measured
LOWFS slopes for two different instrument setups and a comparison with
the focus projection-slopes extracted from the synthetic
reconstruction matrix.  We applied a 200\,nm RMS defocus with the AO
system. This was done in closed loop by the means of a reference
centroid offset to avoid creeping of the woofer or hysteresis
errors. To help visual inspection only neighbouring subapertures are
connected (in contrary to the matrix slopes comparison in
Fig.~\ref{fig:compare_mat}).  We remind the reader that for a pure focus aberration
the slopes are expected to rise linearly in $x$ and to be constant in
$y$ for each subaperture row which is clearly reflected. The
dash-dotted line displays the measured LOWFS slopes for a focus test
wavefront with the apodizer (Apod\_H) inserted (typical coronagraphic
set up) and the simple dashed line measures the same test wavefront
but with the apodizer mask removed (Clear). This verifies that the
discrepancies between these slopes are small. Nevertheless, if they are
completely negligible and can be safely ignored is still a matter of
investigation.


The continuous line in Fig.~\ref{fig:slopes_compare} represents the
values of the focus row from the original synthetic reconstructor
matrix. These projection-slopes are scaled to match the measured
slopes. The discrepancy of the projection-slopes to the measured
slopes arises because of instrumental effects that could not be
simulated and because of the matrix inversion process involving all 14
Zernikes. This demonstrates the advantages of a reconstruction matrix
based on measurements and also reminds us of the trade-off done by the
selection of Zernike modes contained in the matrix. For example, in
our geometry the focus row is significantly modified by the inclusion
of spherical during the inversion process (fitting).

\subsection{Low sensitivity for the spherical mode}
\begin{figure}
   \begin{center}
   \begin{tabular}{c}
   \includegraphics[angle=0,width=.25\textwidth]{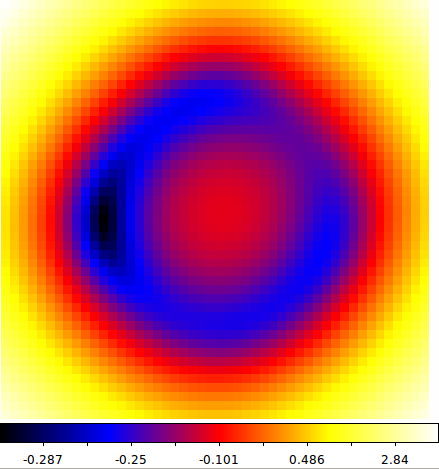}
   \end{tabular}
   \end{center}
   \caption[example] 
   { \label{fig:phase_grid} A 48x48 phase grid as it would be rendered to the AO system, containing the reconstructed phase of a 200 nm RMS focus measured by the LOWFS (H apodizer in). This figure demonstrates the appearance of the spherical mode.}
   \end{figure}

Fig.~\ref{fig:phase_grid} shows a 48x48 phase grid that is generated
for a 200 nm RMS focus wavefront by the CAL interface and which is
rendered to the AO system to offload NCPAs. The phase is purely based
on LOWFS slopes and reconstructed as described in
Sect.~\ref{sect:offload}. An erroneous primary spherical aberration
can be identified which originates from noise amplification or the low
sensitivity of the LOWFS to measure this mode. This is a consequence of
the central obscuration invalidating the central sub-aperture which
contains strongly weighted information to fit a spherical. The best
solution in our case might be to reject this mode completely. The
small but notable discrepancy for circular symmetry could be real
NCPA. If not they are expected to disappear when the original
synthetic reconstruction matrix (used in this case) is replaced by a
measured one.

\subsection{Edge effects seen by the AOWFS}

   \begin{figure}
   \begin{center}
   \begin{tabular}{c}
   \includegraphics[angle=0,width=.5\textwidth]{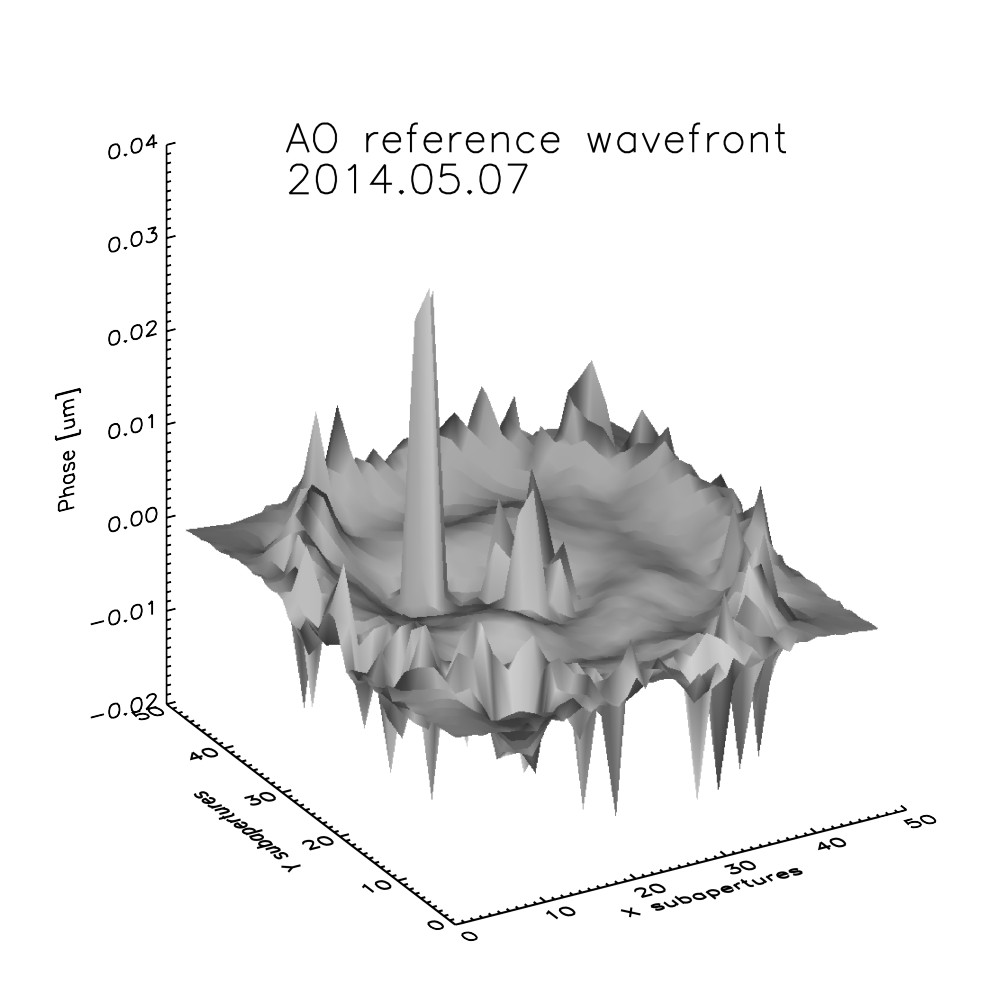}
   \end{tabular}
   \end{center}
   \caption[example] 
   { \label{fig:edge_effects} Wavefront as seen by the AO system used
     for the precision focus calibration. The baseline is subtracted
     and the applied focus is removed. The 20 nm spike is coincident to
     a masked-out dead actuator.}
   \end{figure}

For our test measurement we applied two methods to control the
wavefront and produce a focus with the AO system. The first method
directly commands the woofer in the required shape, working open
loop. This method is expected to be less precise since we can be
impacted by hysteresis and creeping. The second method works in closed
loop and introduces a focus offset via AOWFS reference centroids.  The
introduced focus offset (starting from a flat reference) is expected
to be corrected by the woofer only (low order) and the MEMS should
keep its shape.  
Fig.~\ref{fig:edge_effects} shows a reference wavefront used for LOWFS
focus tests.  Even though the 200\,nm RMS focus is applied it remains
invisible since it is contained in the reference centroids.
Therefore, the displayed wavefront is expected to show mainly noise.
It is interesting to see that the MEMS appears not to remain entirely
unmodified. Handling edge effects on a circular pupil in an optimal
way using a Fourier Controller as GPI does is still subject to ongoing
research.  The ``large'' (20 nm) spike is coincident with a dead
actuator that is masked out.

  \begin{figure}[t!]
   \begin{center}
   \begin{tabular}{c}
   \includegraphics[angle=0,width=1\textwidth]{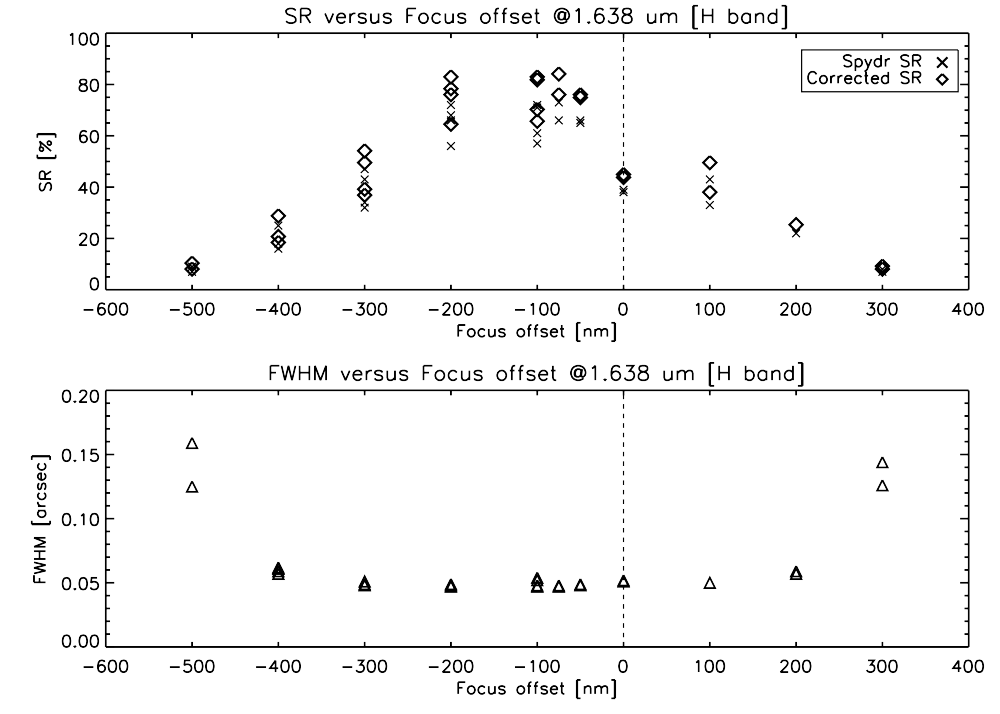}
   \end{tabular}
   \end{center}
   \caption[example] 
   { \label{fig:SR} SR (top) and FWHM (bottom) plotted against the
     defocus applied ; this allows to determine and verify the on-sky
     focus correction. Measured SR from IFS images is displayed as
     crosses, while the diamonds show the same values adjusted by a SR
     correction factor ($\sim$1.15) obtained from a diffraction
     simulation.}
   \end{figure}

Moreover, since March 2014 the AO system was modified to handle the
focus correction in a separate loop and therefore focus is removed
from the slopes before they are fed to the Fourier controller, in
order to increase the performance.  This feature was successfully
commissioned during the May 2014 commissioning run. We also noted a
significantly improved LOWFS reference wavefront using this new AO
focus loop implementation. This improvement is not too surprising
since a Fourier Controller is not optimally suited to handle the very
low order wavefront aberrations (tip/tilt, focus, astigmatism). With
the successful implementation of the separate AO loop to control
focus, we might consider doing the same for astigmatism (similar
slopes as focus apart from sign).

\section{On-sky commissioning tests}
During the last night of May 2014 commissioning run (2014-05-13), we
acquired a key data set to proceed with the LOWFS's focus offload
commissioning. We selected a binary star as target and applied a
series of different focus offsets. The binary target allowed us to
use the coronagraphic mode and having the binary star as Strehl
reference target at the same time. For each of these focus offsets, a
comprehensive set of AO and CAL telemetry data was saved and science
data cubes were taken with GPI's Integral Field Spectrograph (IFS).

From the GPI pipeline\cite{perrin_2014} reduced and wavelength
calibrated data cubes we selected images at $\lambda = 1.638$\,$\mu$m.
This wavelength corresponds to a clean, well-suited frame for SR
measurement in the middle range of the data cube where also the
wavelength solution should be most reliable.  The SR and FWHM values
of the companion were calculated using the Spydr software
package\cite{rigaut_2014} for each focus value. These parameters are
plotted against the focus offset in Fig.~\ref{fig:SR}. The SR is
indicated by crosses and the FWHMs by triangles. The best SR
performance is reached for a focus offset of -150 nm peak to
valley. The off-load of this focus value yields a non-negligible
improvement in the instrument's performance (see
Sect.~\ref{sect:intro}) and should be fully automated during the
course of this year with the commissioning of LOWFS's non-common path
(focus) correction. We remind the reader that this on-sky focus offset
refers to the science detector and not the focal plane mask. At a
future moment, another key experiment for a precision calibration
of the LOWFS will be to perform a knife-edge experiment on-sky. 

The SR ratios are subject to a large error (as indicated by the
dispersion) mainly due to the error of background correction on the
images. Nevertheless, the largest values are around the 80\% mark
(H-band) which corresponds to a residual wavefront error of 120 nm
RMS. This is roughly twice the value then given by our goal error
budget. But these measurements were undertaken under the impact of the
significant vibrations that are currently mitigated. We recently
demonstrated that the mitigation of these vibrations through active
compensation using external hardware or Kalman filtering on the AO
controller side will significantly improve our error budget in respect
to these measurements\cite{poyneer_2014, hartung_2014}.

\begin{figure}
   \begin{center}
   $
   \begin{tabular}{ccc}
   \includegraphics[angle=0,width=0.25\textwidth]{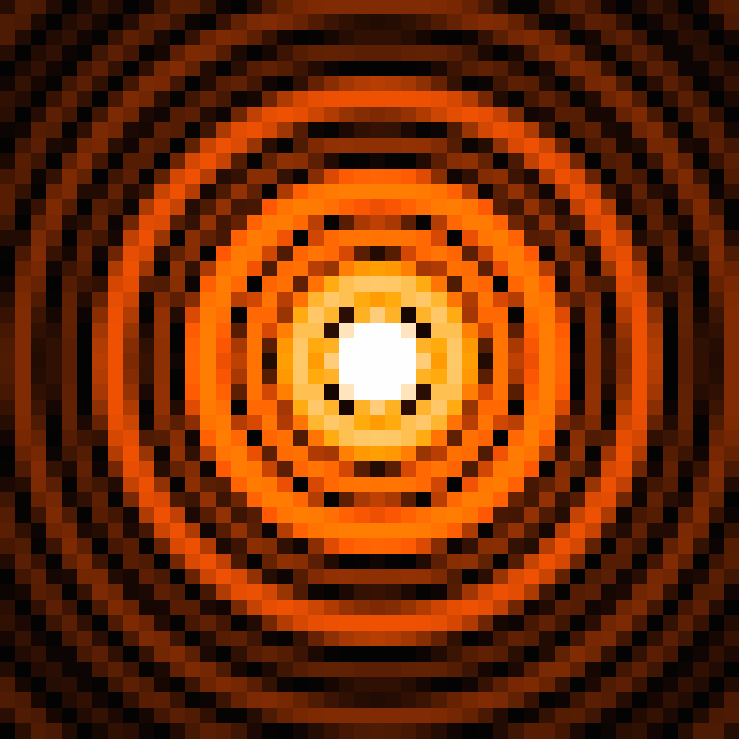} &
   \includegraphics[angle=0,width=0.25\textwidth]{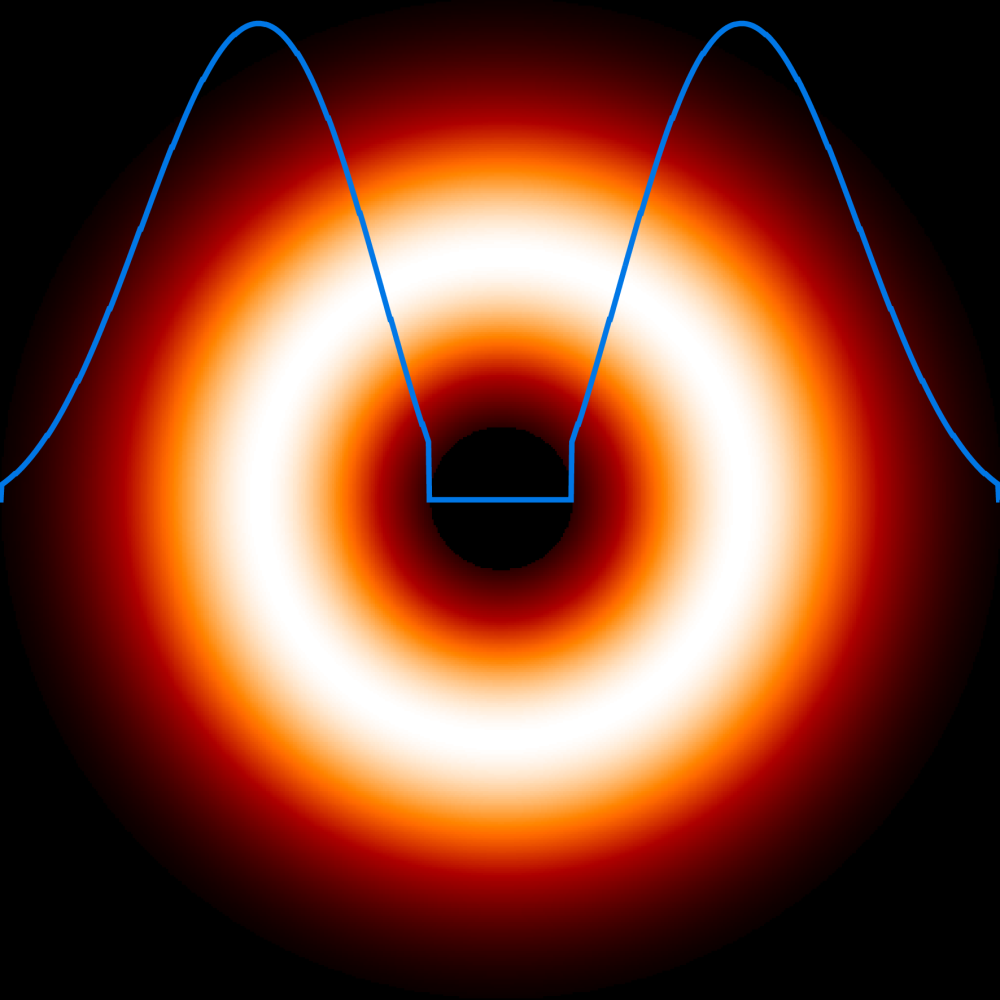} &
   \includegraphics[angle=0,width=0.25\textwidth]{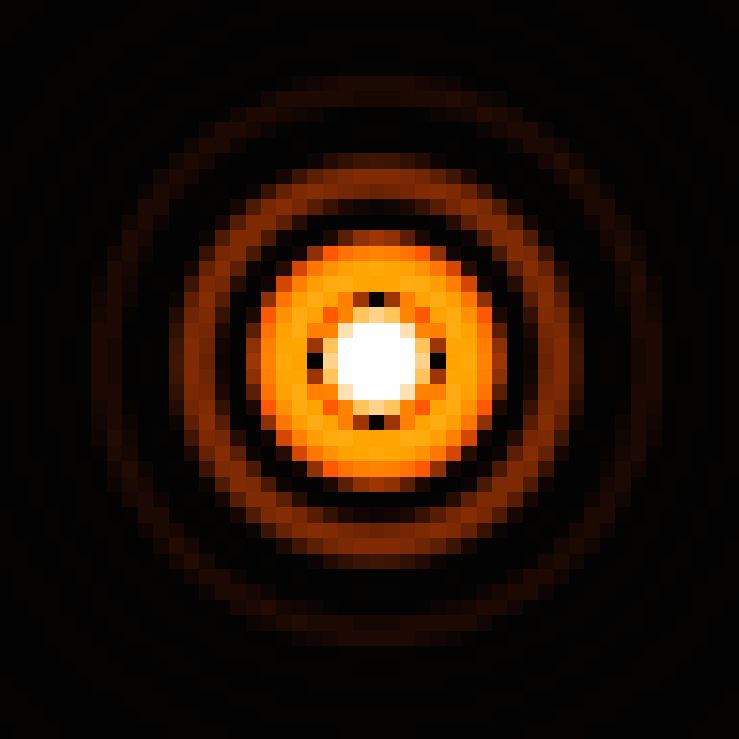} \\
   (a) & (b) & (c) \\
   \end{tabular}
   $
   \end{center}
   \caption[example] 
   { \label{fig:trio} (a) Telescope's simulated PSF (colored in logscale) (b) Apodizer's profile (H-69\_10) ; the apodizer is located in a pupil plane. (c) Apodized telescope's PSF (colored in logscale)}
   \end{figure}

Since the effect of GPI's apodizer profile on the SR is not taken into
account in the Spydr software package, we calculated a correction
factor. We obtained this correction factor by simulating the
effect of the apodizer profile on the PSF:

First, we simulate the diffraction of an incoming homogeneous flat
wavefront on our telescope aperture ($D = 7.7$\,m, central obscuration
= 1.024\,m) and we match the spatial sampling to the IFS science
detector (14\,mas/pixel). In a second step, we ``slide'' the H
apodizer in.  Fig.~\ref{fig:trio} shows how the PSF is reshaped by the
apodizer.

Finally, the SR correction factor is calculated as the intensity ratio
between each set-up's PSF (with and without apodizer), with normalized
light inputs (the apodizer roughly takes out half of the incident
photons). With this approach we obtain a correction factor of
$\sim$1.15 for the SR.


\section{Summary}
The First Light run in November 2013 and the following commissioning
runs have been an impressive success. In April 2014 GPI was already
offered to the community in an Early Science run {\it without} fully
implemented mitigation strategies to cancel
vibrations\cite{hartung_2014} and without the powerful upgrades of the
AO controller\cite{poyneer_2014} (extraction of rotation signal from
centroids of the AO WFS, separate focus AO loop and removal of focus
from the Fourier Controller, and the modal gain optimizer). These
enhancements are currently commissioned and integrated for standard
science operations\cite{rantakyro_2014}. Still, the community was more
than satisfied by the delivered data and already scientific
publications have been produced.

This cleared the road to work on the details for ultimate XAO and high
contrast performance such as optimizing NCPA suppression. This paper
focused on one of the many elements to address this matter. We shared
our challenges and the approach to continue the commissioning of the
CAL unit, starting with the low-order NCPA compensation.  It is a
large effort to achieve the ultimate goals (sense and control the
wavefront with a precision of a few nm) in an ``adverse'' telescope
environment (vibrations, thermal drifts, changing
gravity). Nevertheless, these challenges have been addressed from
different ends\cite{hartung_2014, poyneer_2014, sadakuni_2014} and
enabled the commissioning of the CAL/LOWFS unit. Technical work on the
interface details between the AO system and the CAL is on-going. The
next and last foreseen commissioning run of GPI will take place in
September 2014 and we demonstrated to have all tools in place to
successfully finish the commissioning of NCPA low-order offload. Even
without high-order compensation by the HOWFS commissioned yet this
will be a major step to keep contrast performance optimal and ensure
by itself that GPI can be safely transferred to Gemini science
operations by the end of 2014.

\acknowledgments     
The Gemini Observatory is operated by the Association of Universities
for Research in Astronomy, Inc., under a cooperative agreement with
the NSF on behalf of the Gemini partnership: the National Science
Foundation (United States), the National Research Council (Canada),
CONICYT (Chile), the Australian Research Council (Australia),
Minist\'{e}rio da Ci\^{e}ncia, Tecnologia e Inova\c{c}\~{a}o (Brazil)
and Ministerio de Ciencia, Tecnolog\'{i}a e Innovaci\'{o}n Productiva
(Argentina).


\bibliographystyle{spiebib}   

\end{document}